\documentclass{aa}
\usepackage{epsfig}
\usepackage{amsmath}
\usepackage{txfonts}
\usepackage{natbib}
\usepackage{bm}

\begin{document}

\title{Damping of coronal oscillations in self-consistent 3D radiative MHD simulations of the solar atmosphere}
\authorrunning{Kohutova et al.}
\titlerunning{Damping of coronal oscillations in 3D MHD simulations}
\author{P. Kohutova\inst{1,2}, P. Antolin\inst{3}, M. Szydlarski\inst{1,2}, and M. Carlsson\inst{1,2}}
\institute{Rosseland Centre for Solar Physics, University of Oslo, P.O. Box 1029, Blindern, NO-0315 Oslo, Norway\\
\email{petra.kohutova@astro.uio.no}
\and
Institute of Theoretical Astrophysics, University of Oslo, P.O. Box 1029, Blindern, NO-0315 Oslo, Norway
\and
Department of Mathematics, Physics and Electrical Engineering, Northumbria University, Newcastle Upon Tyne NE1 8ST, UK}
\date{Received; accepted}

\abstract
{Oscillations are abundant in the solar corona. Coronal loop oscillations are typically studied using highly idealised models of magnetic flux tubes. In order to improve our understanding of coronal oscillations, it is necessary to consider the effect of realistic magnetic field topology and density structuring.} 
{We analyse the damping of coronal oscillations using a self-consistent 3D radiation-MHD simulation of the solar atmosphere spanning from the convection zone into the corona, the associated oscillation dissipation and heating, and finally the physical processes responsible for the damping and dissipation. The simulated corona formed in such a model does not depend on any prior assumptions about the shape of the coronal loops.}
{We analyse the evolution of a bundle of magnetic loops by magnetic field tracing.}
{We find that the bundle of magnetic loops shows damped transverse oscillations in response to perturbations in two separate instances with oscillation periods of 177 s and 191 s, velocity amplitudes of 10 km s$^{-1}$ and 16 km s$^{-1}$ and damping times of 176 s and 198 s, respectively. The coronal oscillations lead to the development of velocity shear in the simulated corona resulting in the formation of vortices seen in the velocity field caused by the Kelvin-Helmholtz instability, contributing to the damping and dissipation of the transverse oscillations.}
{The oscillation parameters and evolution observed are in line with the values typically seen in observations of coronal loop oscillations. The dynamic evolution of the coronal loop bundle suggests the models of monolithic and static coronal loops with constant lengths might need to be re-evaluated by relaxing the assumption of highly idealised waveguides.}


\maketitle

\section{Introduction}
Coronal structures act as waveguides for a variety of MHD oscillation modes \citep{nakariakov_2005, nakariakov_2020}. There is extensive observational evidence that transverse oscillations are ubiquitous in the solar corona, in both closed coronal loops and open coronal structures \citep{tomczyk_2007, mcintosh_2011, anfinogentov_2015}. The transverse oscillations commonly observed in coronal loops are identified as standing kink modes \citep{aschwanden_1999, nakariakov_1999}. Coronal oscillations carry magnetic energy, which is deposited through the oscillation damping and dissipation. Commonly proposed physical mechanisms responsible for the oscillation damping and dissipation include resonant absorption \citep{hollweg_1988, goossens_2002}, phase mixing \citep{heyvaerts_1983} and Kelvin-Helmholtz instability \citep{terradas_2008, antolin_2014}. Such mechanisms have been mostly analysed using simplified models of coronal loops as straight magnetic flux-tubes clearly distinct from the surroundings. There are several advantages to this approach: It is straightforward to isolate individual effects and processes linked to wave evolution as well as to have full control over the parameters of the modelled loop. They also allow for very high spatial resolution, necessary for modelling of certain physical processes such as resonant absorption \citep{doorsselaere_2004}. Such models, however, also require making several assumptions about the shape, density structure and morphology of coronal loops which might be too idealised. 

The solar corona is in fact a dynamic environment with complex density structuring and the coronal magnetic field is continuously changing and evolving. Models of static and idealised coronal loops therefore neglect this evolution. In order to account for realistic magnetic field configurations and density structuring in the solar corona, a more self-consistent approach to modelling the evolution of coronal structures is necessary. This can be achieved by taking advantage of realistic convection-zone-to-corona models \citep[e.g.][]{carlsson_2016, cheung_2019, kohutova_2021, breu_2022}. The evolution of the corona in such models is self-consistently driven by the dynamics of the lower solar atmosphere. This type of simulations therefore reflects the dynamic and continuously evolving nature of the solar corona. 

One common feature of such simulations is the lack of clearly-identifiable coronal loops with well-defined boundaries. In the realistic solar simulations the actual 3D structure of coronal features with increased emissivity in the optically thin coronal emission lines is much more complex, despite appearing as thin, well-defined loops in forward-modelled emission due to line-of-sight effects. Such structure of the corona is referred to as 'the coronal veil', and has been described using a self-consistent MURaM simulation spanning from the convection zone into the corona \citep{malanushenko_2022}. Different codes capable of producing self-consistent convection-zone-to-corona models, including MURaM \citep{rempel_2016} and Bifrost \citep{gudiksen_2011}, seem to reproduce the 'coronal veil' structure of the solar corona. The question remains how well does the simulated corona, which forms in these models, represent the real solar corona. On one hand, the coronal veil model can explain certain properties of the coronal loops seen in 
Extreme Ultraviolet (EUV) observations such as loop cross-sections that appear to be constant with height \citep{aschwanden_2005}, which are difficult to explain otherwise; we note that alternative explanations include a specific density and temperature distribution across the magnetic field \citep{peter_2012} and the presence of magnetic twist \citep{li_2020}. It would however also bring the applicability of coronal loop oscillation models into question, as most rely on the assumption of coronal loops being well represented by idealised magnetic cylinders. 
On the other hand, the existence and omnipresence of transverse MHD waves such as kink waves directly points to the existence of structures acting as waveguides that lead to collective behaviour in the coronal volume. Also, the occurrence of coronal rain showers \citep{sahin_2022} and long period intensity-pulsations \citep{froment_2015}, ubiquitous over active regions, indicate the existence of coronal entities (that we refer as loops or loop bundles) with similar thermodynamic behaviour.

\begin{figure*}
	\includegraphics[width=43pc]{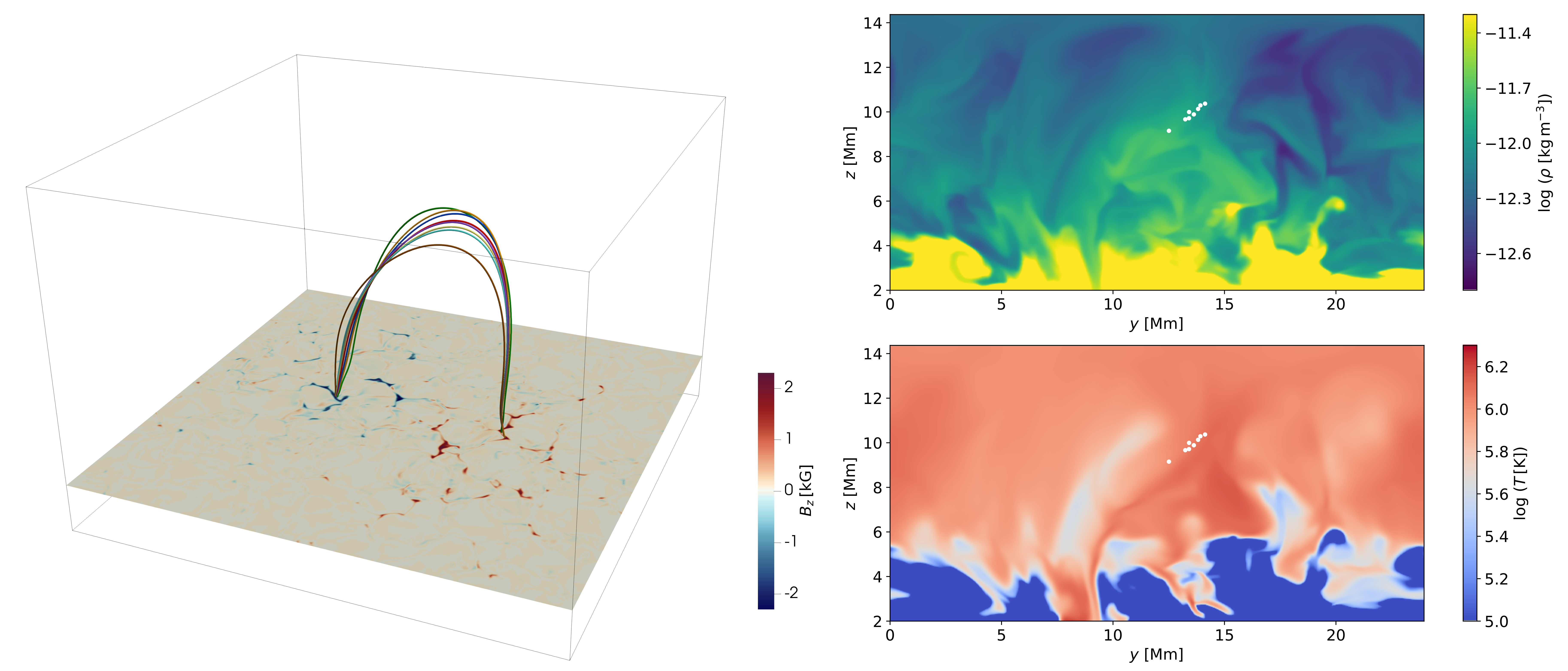}
	\caption{Left: Bundle of magnetic loops and the photospheric magnetic field at $t=880$ s.Top right: The cut across the density structure at $x=13$ Mm intersecting the apex of the magnetic bundle. Bottom right: The cut across the temperature structure at $x=13$ Mm. The white points mark the positions at which the individual loops in the bundle intersect the cuts. An animation of this figure is available online.}
	\label{fig:context_full}
\end{figure*}

Most models for wave dissipation rely on modelling coronal loops as straight magnetic flux-tubes. The mechanism of resonant absorption depends on conversion of a global kink mode into local azimuthal Alfvén modes in an inhomogeneous layer with an Alfvén speed gradient at the boundary of a cylindrical coronal loop \citep{goossens_2002, pascoe_2010, howson_2019}. Similarly, the development of Kelvin-Helmholtz Instability vortices at the boundary of a transversely oscillating loop has been mostly studied using models of cylindrical loops \citep[e.g.][]{antolin_2014, karampelas_2017, howson_2017, howson_2019} with a distinct density or magnetic interface separating them from the background plasma.

The main shortcoming of these models is that the observational evidence for the proposed damping and dissipation mechanisms is largely inconclusive. Signatures reminiscent of resonant absorption have been observed in oscillating prominence threads \citep{antolin_2015, okamoto_2015} and in transversely oscillating spicules \citep{antolin_2018}. To the best of our knowledge, these are the only observational evidence of such mechanisms to date. Models for damping of coronal oscillations would therefore benefit from extending into more realistic setups. Numerical studies using setups which do not correspond to magnetic cylinders were done for both oscillations in the chromosphere \citep{leenaarts_2015, khomenko_2012} and in the corona \citep{matsumoto_2018}, the latter two however lack the self-consistent treatment of the lower solar atmosphere driving the corona.

In this work we therefore focus on analysing coronal oscillations in a more advanced numerical setup. We exploit the potential of convection-zone-to-corona simulations with the radiation-MHD code Bifrost for coronal studies. We have previously shown that coronal oscillations are abundant in self-consistent convection-zone-to-corona simulations and that the detected oscillation modes and regimes match those seen in solar observations \citep{kohutova_2021}. Here we analyse the evolution of an oscillating bundle of magnetic loops which forms in the simulation and focus on the oscillation damping. Despite the complex evolution of the bundle, variable length etc. the individual magnetic field lines contained in the bundle are found to exhibit collective evolution during extended periods of time, including collective oscillations triggered by impulsive events and subsequent oscillation damping. 

The manuscript is structured as follows. Section \ref{section:model} describes the numerical model. Section \ref{section:evolution} describes the methods used for the analysis of the evolution of a bundle of magnetic loops and of the corresponding oscillatory behaviour seen in the simulation. In Section \ref{section:oscillations} we focus on the damping of the oscillating loops and on the physical mechanism responsible for the damping. In Section \ref{section:discussion} we discuss the results and implications for the coronal loop models. We summarize our conclusions in Section \ref{section:conclusions}.

\begin{figure*}
	\includegraphics[width=43pc]{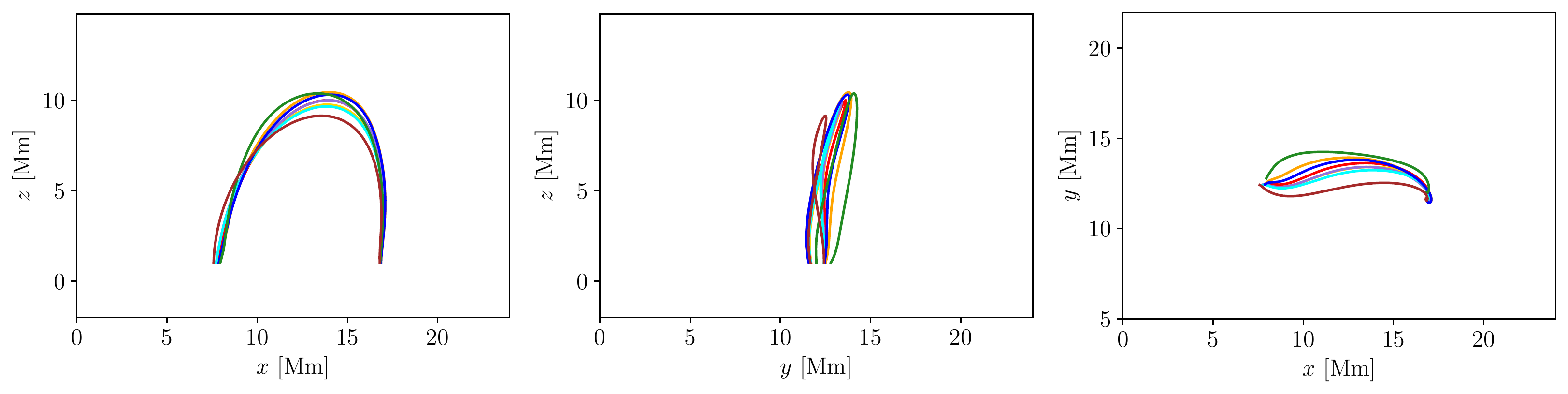}
	\caption{A projected view of the magnetic bundle in the $xz$-plane (left), $yz$-plane (centre) and $xy$-plane (right) at $t=880$ s. An animation of this figure is available online.}
	\label{fig:projection}
\end{figure*}

\section{Numerical model} 
\label{section:model}
We analyse the evolution, oscillatory behaviour and damping of a bundle of magnetic loops in the numerical simulation of a magnetically enhanced network spanning from the upper convection zone to the corona using the Bifrost code \citep{gudiksen_2011}. This simulation corresponds to the extended run of the public Bifrost simulation of the enhanced network \citep{carlsson_2016}. The simulation subset with 2000 s duration analysed in this work covers the time range from t = 100 s to t = 2100 s of the extended run of the enhanced network simulation run from the last snapshot of the public simulation, while having the same physical setup. 

Bifrost is a 3D radiation-MHD code which solves the resistive MHD equations and includes radiative transfer with scattering in the photosphere and low chromosphere, and parametrised radiative losses and heating in the upper chromosphere, transition region and corona. The effects of field-aligned thermal conduction and the non-equilibrium ionisation of hydrogen in the equation of state are included in the simulation.

The physical size of the simulation grid is 24 $\times$ 24 $\times$ 16.8 Mm and the grid resolution is 504 $\times$ 504 $\times$ 496. The grid spans from 2.4 Mm below the photosphere to 14.4 Mm in the corona. The photosphere is located at $z=0$ surface and corresponds to the (approximate) height where the optical depth $\tau_{500}$ is equal to unity. The grid spacing is 48 km and uniform in the $x$ and $y$ direction, while in the $z$ direction it varies from 19 km to 98 km in order to resolve steep density and temperature gradients in the lower solar atmosphere.

The simulation domain uses periodic boundaries in the $x$ and $y$ directions and open boundaries in the $z$ direction. The top boundary uses characteristic boundaries which transmit disturbances with minimal reflection \citep{gudiksen_2011}. At the bottom boundary the flows are let through and the magnetic field is passively advected without introducing any additional magnetic field into the domain. The average horizontal pressure is driven towards a constant value with a characteristic timescale of 100 s, creating a pressure node at the bottom boundary. This leads to acoustic wave reflection resembling the refraction of waves in the deeper solar atmosphere, resulting in global radial box oscillations with a period of 450 s, which are a simulation counterpart of solar p-modes \citep{stein_2001, carlsson_2016}.

The photospheric magnetic field is concentrated in two patches of opposite polarity and has an average unsigned value of about 50 G (Fig. \ref{fig:context_full}). The dipolar structure of the  magnetic field creates several magnetic loops in the simulated corona. The convective motions in the lower solar atmosphere lead to magnetic field braiding. The Ohmic and viscous heating associated with the braiding together maintain high temperatures in the chromosphere and corona. An artificial heating term is employed for plasma with temperatures below 2500 K, in order to prevent the temperature from dropping too low in regions that are rapidly expanding. Only few isolated regions in the simulation domain are affected by this, and the heating in the vast majority of the domain is self-consistent. Contributing to the heating are small-scale reconnection events which heat the plasma through a combination of direct Ohmic dissipation and by inducing shear flows which are then converted into heat by viscous dissipation as well as dissipating oscillations. In order to ensure numerical stability the code employs a diffusive operator; this consists of a small global diffusion term as well as of a directionally-dependent hyper diffusion component which enhances the diffusion locally. Further details of the numerical setup can be found in e.g. \citet{carlsson_2016, kohutova_2020}.

\section{Evolution of a magnetic loop bundle}
\label{section:evolution}
The corona in the simulation is filled with closed magnetic loops which can extend up to heights of $10-14$ Mm. The density structure of the simulated corona is complex and there are several structures with enhanced densities compared to the surroundings. Most of the overdense structures are filled by chromospheric evaporation in response to heating \citep{kohutova_2020}. A cut across the simulation domain shows a lack of loops with clearly defined cross-sections in the density or temperature structure (Fig. \ref{fig:context_full}).

The magnetic field configuration in the simulation domain is driven by the dynamics of the lower solar atmosphere and the footpoints of coronal structures are shuffled and dragged around by the convective motions. The magnetic loops in the corona are therefore continuously evolving and undergoing complex motions including sideways displacement, oscillatory and torsional motion and vertical expansion/contraction. 

We focus on the evolution of a bundle of magnetic loops located in the centre of the simulation domain shown in Fig. \ref{fig:context_full} reaching a height of around 10 Mm. Due to the evolving nature of the magnetic field in the simulated corona, in order to obtain the evolution of such a bundle in three dimensions it is necessary to trace the evolution of the corresponding magnetic field lines in the bundle through both time and space. To do this, we use a field-tracing method described in \citet{leenaarts_2015} and \citet{kohutova_2021}. 

A magnetic field line is defined as a curve in a 3D space $\vec{r}(s)$ parametrised by the arc length along the curve $s$, for which $\mathrm{d} \vec{r}/ \mathrm{d} s = \vec{B} / |\vec{B}|$ and is therefore a representation of magnetic connectivity of the coronal plasma. We trace the evolution of magnetic field lines by inserting seed points into the simulation domain at the apex of the magnetic bundle. Using the velocity at the seed point position the seed points are then passively advected forward and backward in time. The magnetic field is then traced through the instantaneous seed point position in order to determine the spatial coordinates of the traced field line at every time-step. The accuracy of this method is given by the size of the time-step between two successive simulation snapshots (i.e., 1 second in this case). We find that the method works well for 10 s step-size and there are no major differences in field-line evolution between 1 s and 10 s time-step size. The 10 s time-step size is therefore used for the field line tracing in the following analysis. We note that this approach requires that the evolution is smooth and there are no large-amplitude velocity variations occurring on timescales shorter than the size of the time-step. Similarly, in the instances where magnetic reconnection occurs, this approach fails and the tracing leads to a jump in the field-line evolution. This is, however, not the case in the coronal part of the magnetic bundle during the analysed time-period. We note there are 2 instances of rapid transverse displacement occuring close to the foootpoints of the traced loops at t = 830 s and t = 1100 s caused either by an external perturbation or change of magnetic connectivity in the lower solar atmosphere. These however do not lead to discontinuities in the evolution of physical quantities in the coronal part of the analysed loops.

The magnetic bundle is shown in different projections in Fig. \ref{fig:projection} with the bundle evolution shown in the online animation. The individual magnetic loops display a large degree of collective behaviour and the magnetic bundle behaves as a coherent structure during most of the duration of the simulation. The footpoints of the magnetic bundle are not static, and the bundle is continuously changing and evolving. The lengths of the individual loops in the magnetic bundle change significantly over the duration of the simulation, sometimes on a timescale of minutes.

\begin{figure}
	\includegraphics[width=21pc]{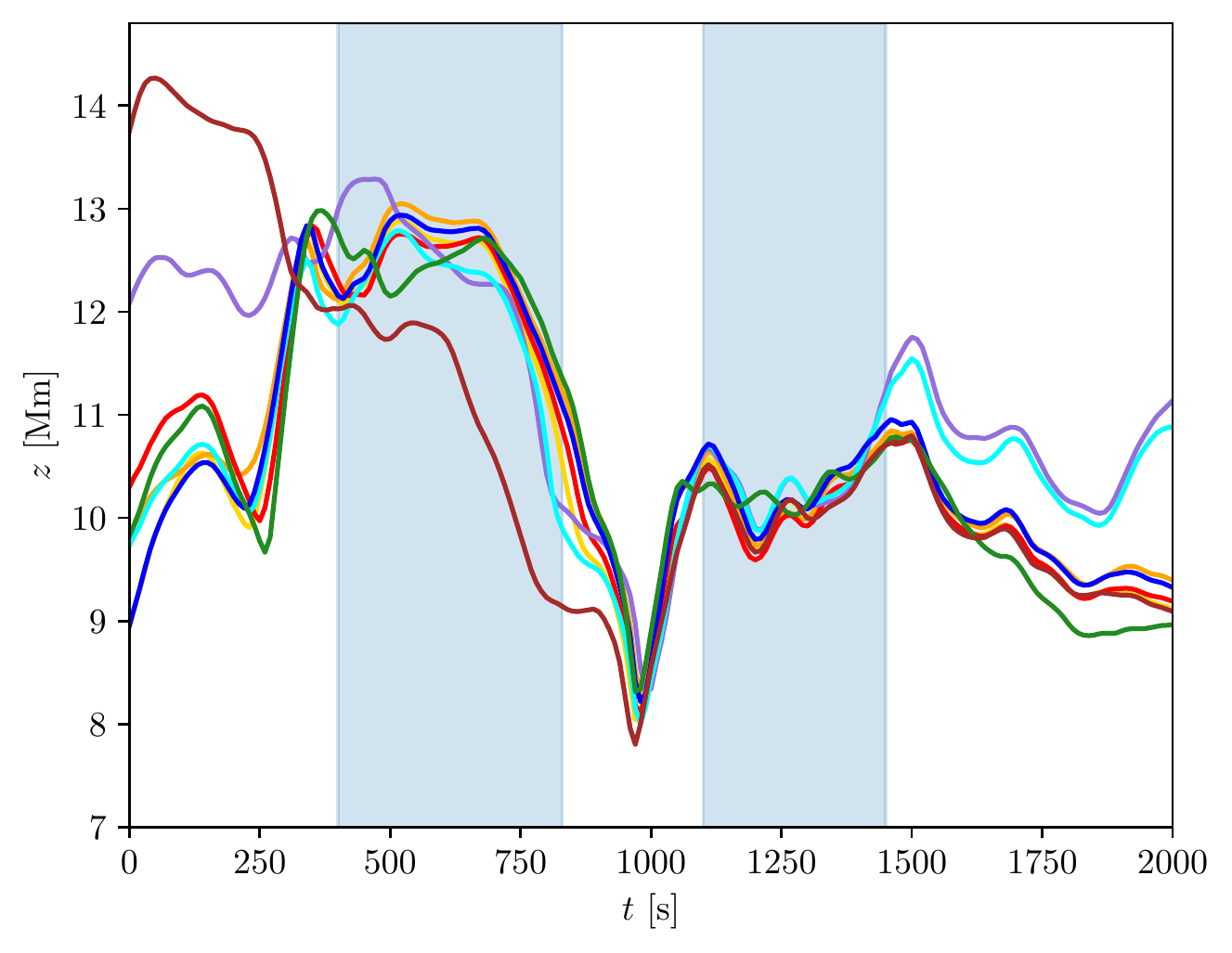}
	\caption{The evolution of the height of the individual loops in the bundle at $x=13$ Mm as a function of time. Blue regions indicate the two instances of damped oscillatory motion in the vertical direction.}
	\label{fig:evolution}
\end{figure}

\begin{figure}
	\includegraphics[width=21pc]{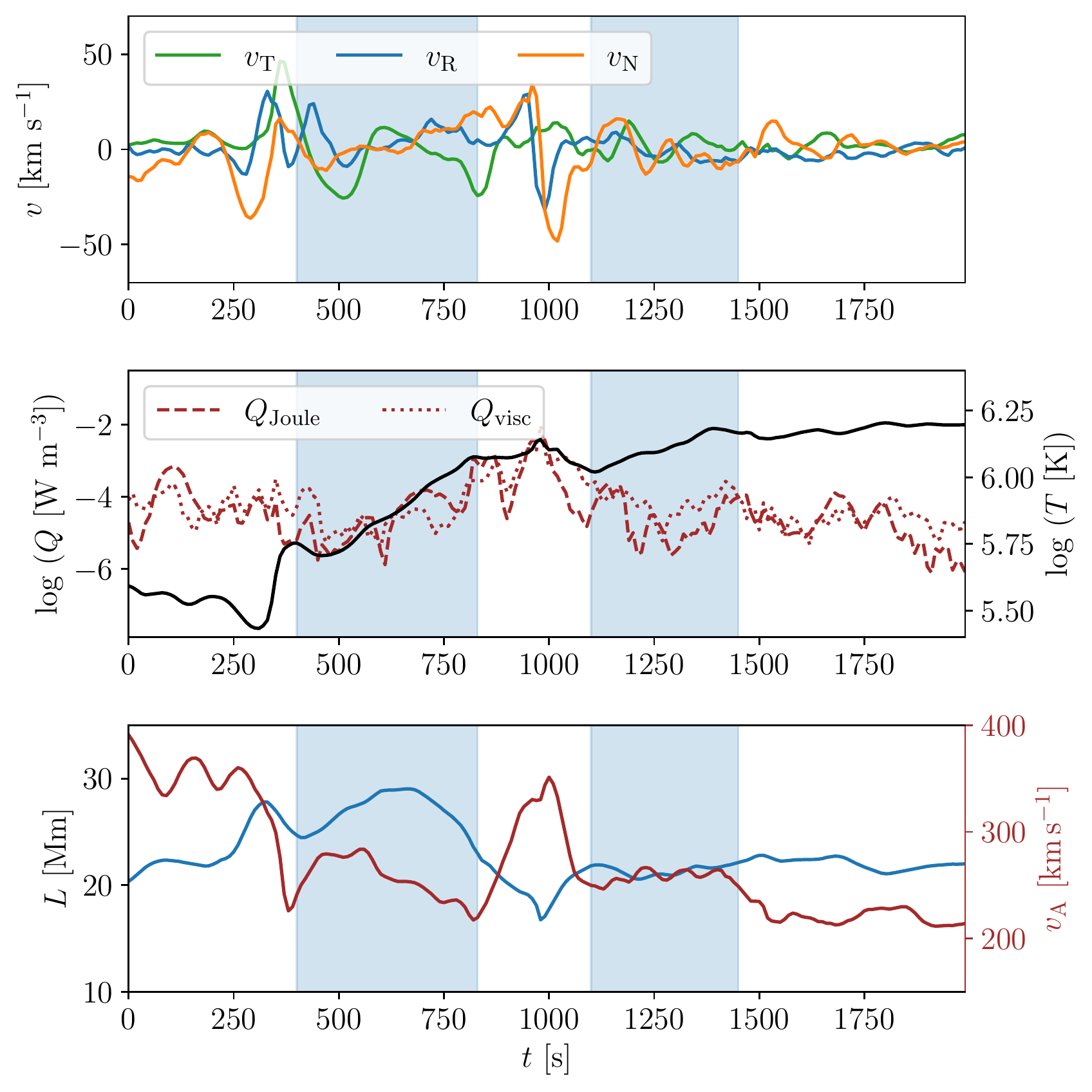}
	\caption{Top: The evolution of the longitudinal (green), normal (orange) and the binormal velocity component (blue) at the apex of the bundle. Middle: The evolution of the Joule volumetric heating rate (red dashed), the viscous volumetric heating rate (red dotted) and the temperature (black) at the apex of the bundle. Bottom: The evolution of the average length (blue) and the Alfvén velocity at the apex of the loops in the bundle. Blue regions indicate the time-range corresponding to the two oscillations.}
	\label{fig:fullt}
\end{figure}

\begin{figure*}
	\includegraphics[width=43pc]{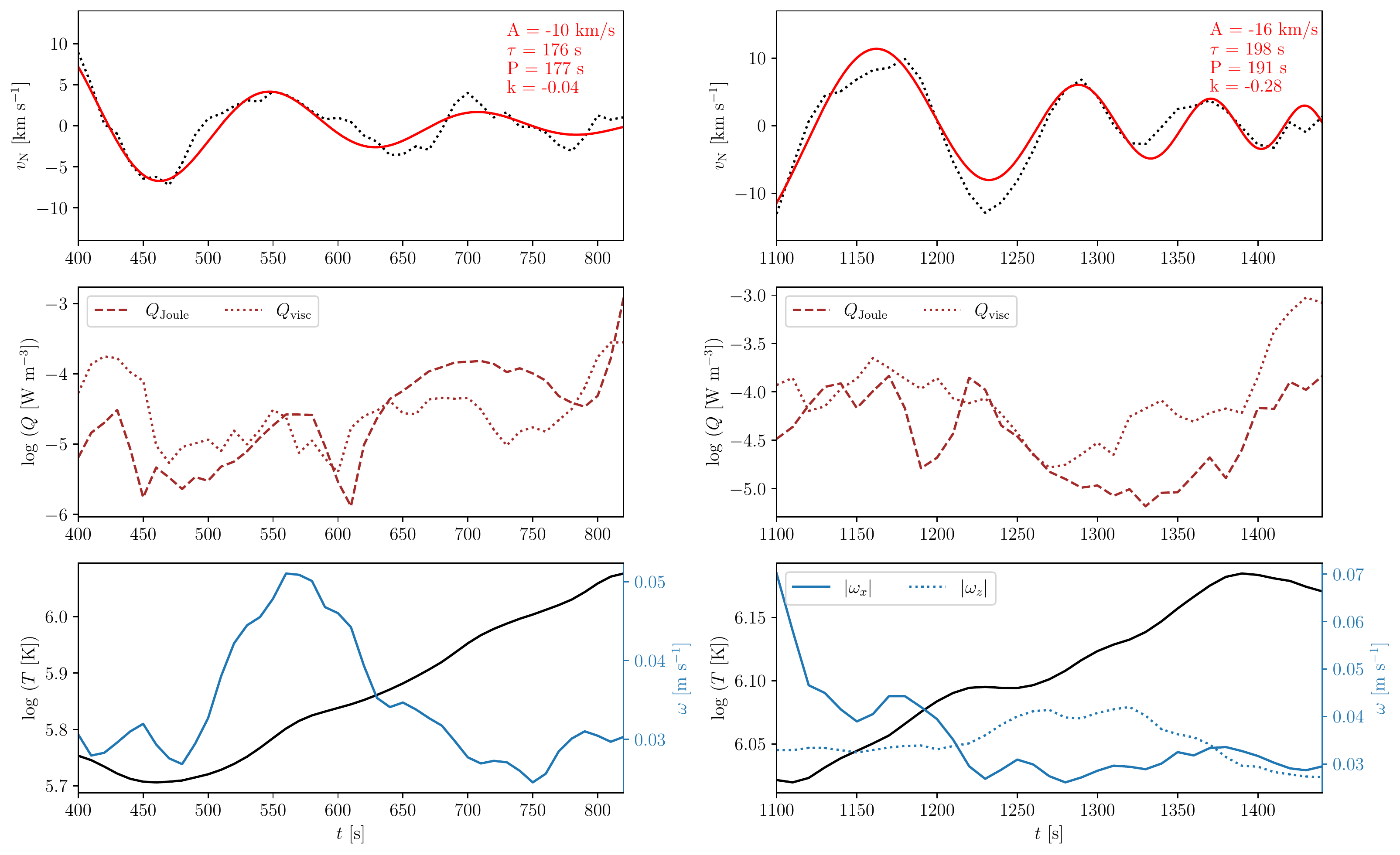}
	\caption{Top: The evolution of the detrended normal velocity component at the bundle apex averaged over the oscillating fieldlines in the bundle. Best-fit is shown in red. Middle: The evolution of the Joule (dashed line) and viscous volumetric heating rate (dotted line) at the bundle apex. Bottom: Temperature (black), the $x$-component of the vorticity averaged in the vicinity of the bundle apex (solid blue line) and the $z$-component of the vorticity averaged in the vicinity of the right bundle footpoint (dotted blue line). The time intervals shown correspond to the time ranges highlighted in blue in Figs. \ref{fig:evolution} and \ref{fig:fullt}.}
	\label{fig:oscillations}
\end{figure*}

\begin{figure*}
	\includegraphics[width=43pc]{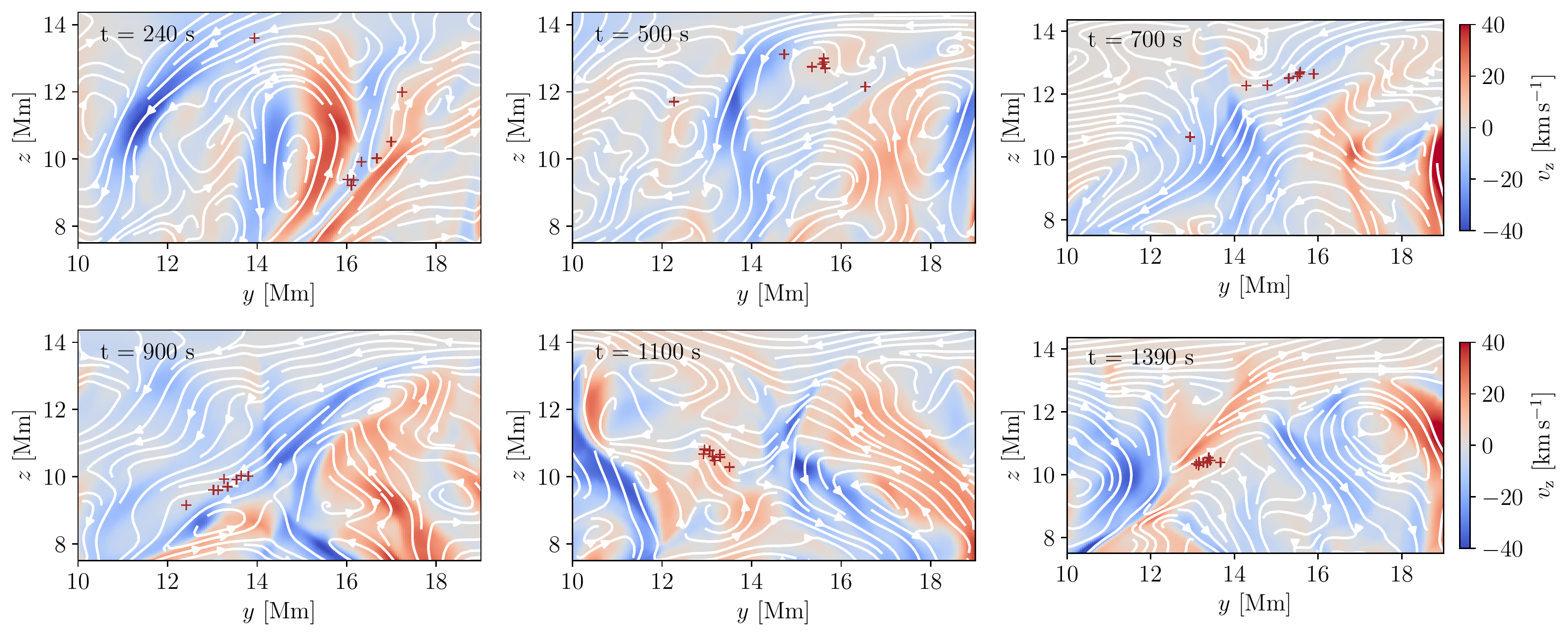}
	\caption{Top: The $v_z$ velocity component and the corresponding velocity field streamlines in cuts parallel to the $y-z$ plane at $x = 13$ Mm shown before the oscillation onset (left), during the oscillation (middle) and at the end of the oscillation (right). Bottom: Same as above but for oscillation 2. The red cross markers correspond to the positions at which the individual loops in the bundle intersect the cut. The locations of the enhanced shear in the $v_z$ velocity component correspond to strong counter-directional flows. In both cases several vortices develop in the vicinity of oscillating structures.}
	\label{fig:vortices}
\end{figure*}

\begin{figure*}
	\includegraphics[width=43pc]{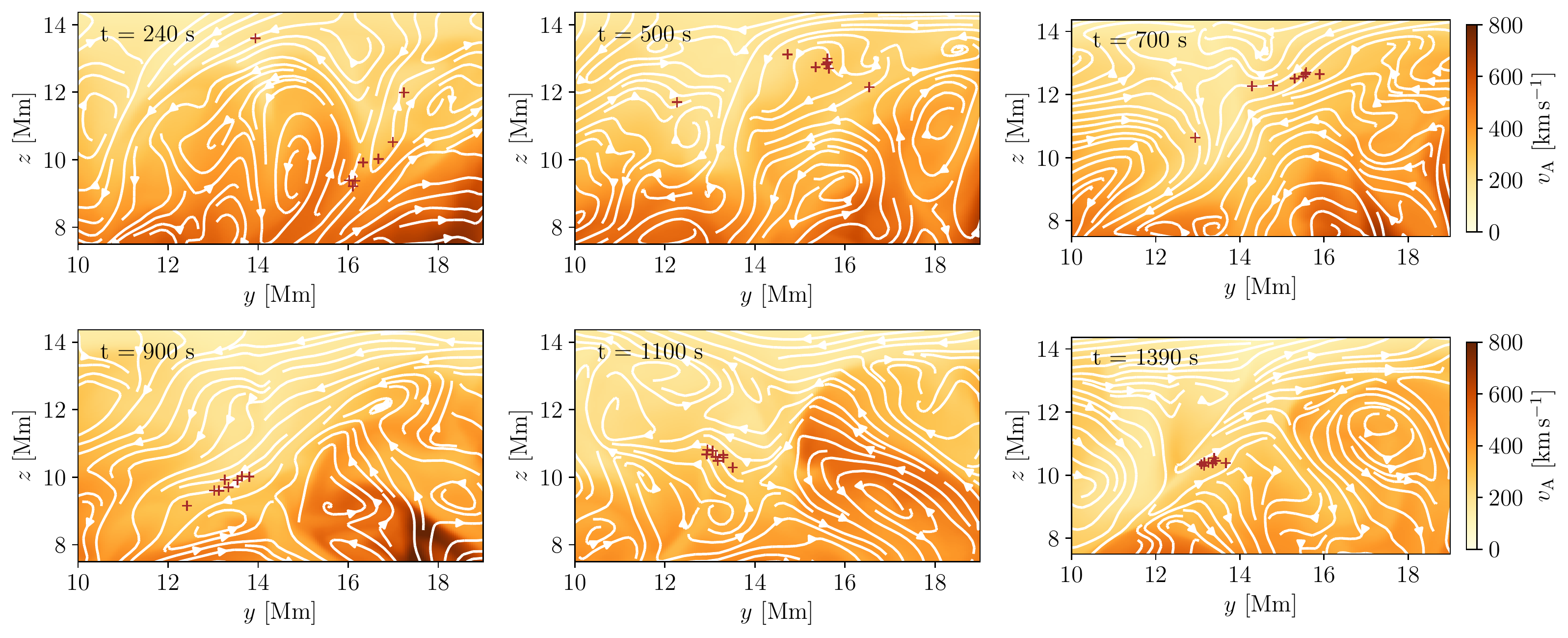}
	\caption{Top: The Alfvén speed and the corresponding velocity field streamlines in cuts parallel to the $y-z$ plane at $x = 13$ Mm shown before the oscillation onset (left), during the oscillation (middle) and at the end of the oscillation (right). Bottom: Same as above but for oscillation 2. The red cross markers correspond to the positions at which the individual loops in the bundle intersect the cut.}
	\label{fig:vortices_va}
\end{figure*}

\section{Oscillations and damping}
\label{section:oscillations}
 Full spatial coordinates of the magnetic loops in the bundle at any point in time enable us to obtain the evolution of the physical quantities along the loops, including both scalar and vector quantities. To aid the oscillation analysis, we decompose the velocity vector into three velocity components relative to the direction of the magnetic field, $v_{\mathrm{T}}$, $v_{\mathrm{N}}$ and $v_{\mathrm{R}}$. The longitudinal velocity $v_{\mathrm{T}} = \vec{v} \cdot \vec{T}$ corresponds to the velocity component aligned with the tangent vector of the magnetic field line given by $\vec{T} = \vec{B}/|\vec{B}|$. The normal velocity $v_{\mathrm{N}} = \vec{v} \cdot \vec{N}$ corresponds to the velocity component along the normal vector of the field line given by $\vec{N} =  \frac{\mathrm{d}\vec{T}}{\mathrm{d}s} / |\frac{\mathrm{d}\vec{T}}{\mathrm{d}s}| $. In the case of a closed magnetic loop, $v_{\mathrm{N}}$ represents the motion in the plane of the loop and perpendicular to the loop tangent. Finally the third velocity component aligned with the binormal vector is given by $v_{\mathrm{R}} =  \vec{v} \cdot \vec{R} $ where $\vec{R} = \vec{T} \times \vec{N}$. In the case of a closed magnetic loop, $v_{\mathrm{R}}$ corresponds to transverse motion perpendicular to the plane of the loop and to the loop tangent. Unit vectors $\vec{T}$, $\vec{N}$, and $\vec{R}$ together form an orthogonal coordinate system known as the Frenet frame of reference. Such coordinate system is well-suited to analysing oscillations in complex 3D magnetic field geometries \citep[e.g.,][]{carlsson_bogdan_2006, felipe_2012, leenaarts_2015, gonzalez_2019, kohutova_2021}.

The collective behaviour of the loop bundle is apparent from Fig. \ref{fig:evolution}, which shows the evolution of the height of the individual magnetic loops in the bundle at $x = 13$ Mm. The evolution shows two clear instances of oscillatory behaviour, starting at $t = 400$ s and $t = 1100$ s and lasting about 400 s in both cases, as indicated marked by blue regions in Figs. \ref{fig:evolution} and \ref{fig:fullt}. The oscillations are damped and occur in a plane perpendicular to the bundle axis. From the evolution shown in animations of Figs. \ref{fig:context_full} and \ref{fig:projection} it is clear that these oscillations correspond to a transverse standing mode of the bundle. The point of maximum oscillation displacement and hence maximum oscillation velocity amplitude lies close to the apex of the bundle with the bundle footpoints acting as nodes of the standing oscillation. The oscillations are triggered by impulsive events in the corona associated with a peak in the Joule heating and subsequent large-scale displacement of the analysed bundle.

We analyse the evolution of the $v_{\mathrm{R}}$ and $v_{\mathrm{N}}$ components by averaging the velocity components over the oscillating loops in the bundle at the loop apex, which we define as the point halfway between the two footpoints of the loop (Fig \ref{fig:fullt}). The oscillations are most clearly detectable in the  $v_{\mathrm{N}}$ component. In both cases the oscillations follow large scale perturbations of the magnetic bundle.

To remove the large-scale trends in the $v_{\mathrm{N}}$ evolution corresponding to a bulk motion of the bundle, the oscillation time-series are detrended by subtracting a best-fit second-degree polynomial. The time-series are then fitted with the function 
\begin{equation}
v(t) = v_0 \exp{\biggl (-\frac{t}{\tau}\biggr )}\sin\biggl (\frac{2 \pi t}{P+kt} - \Phi\biggr )
\end{equation}
which includes both the oscillation damping and linear change in the oscillation period (Fig. \ref{fig:oscillations}). Here $v_0$ corresponds to the velocity amplitude, $\tau$ is the decay time, $P$ is the oscillation period, $k$ is a parameter controlling the linear change in period and $\Phi$ is the phase. We find that the initial velocity amplitudes are 10 km/s and 16 km/s, these are damped with decay times of 176 s and 198 s respectively, corresponding to roughly 3 oscillation periods being detectable before the oscillation decays. The periods for the first and second oscillation are 177 s and 191 s respectively. The oscillation periods in the both cases decrease over the duration of the oscillation; by approximately 15 s for oscillation 1 and 100 s for oscillation 2. The Joule and viscous volumetric heating rates shown in Fig. \ref{fig:oscillations} are also averaged over the oscillating loops at the apex bundle. Over the duration of oscillations the heating rates at the apex of the bundle are variable, but overall have an increasing trend during the later stages of both oscillations. Similarly, the average temperature at the bundle apex increases by around 0.5 MK over the duration of the oscillation in both cases. Finally, we show the evolution of the absolute value of the $x$-component of the vorticity $\omega_x$ in the $y-z$ plane at $x=13$ Mm, averaged over the area surrounding the bundle apex. The $\omega_x$-component is chosen because the axis of the loop at the apex is roughly aligned with the $x$-axis and the oscillatory motion mostly occurs in the plane perpendicular to the loop axis. During the first oscillation, the vorticity in the surroundings of the oscillating bundle increases to reach the maximum value around 170 s after the oscillation onset, followed by a gradual decrease to the pre-oscillation values. In the later case, the vorticity in the vicinity of the bundle apex decreases during the entire oscillation duration. For the second oscillation we also calculate the evolution of the $\omega_z$ vorticity component at the right footpoint at the height of 3 Mm, as this undergoes strong transverse displacement. Similar vorticity peak as at the apex position of the first oscillation can be seen in the evolution here.

To further understand the evolution of the vorticity close to the oscillating bundle, we show the velocity field given by $v_y \vec{e_y} + v_z \vec{e_z}$ in the $y-z$ plane at $x=13$ Mm which is nearly perpendicular to the bundle axis and in the plane of the bundle oscillations (Fig \ref{fig:vortices}). We calculate the velocity field before the onset of the oscillation, during the oscillation and at the end of the oscillation. The shear flows are abundant in the vicinity of the bundle, this shows as counter-directional large magnitude flows in the $v_z$ component of the velocity. We find that the velocity shear is strongest before and during the oscillation. Associated with this is the development of several vortices visible in the velocity streamlines in the close proximity of the oscillating structures. The velocity shear in this region weakens and the vortices mostly disappear once the oscillations have decayed. The vortices visible in the velocity streamlines are always located in the regions with strong velocity shear, suggesting they originate due to the development of the Kelvin-Helmholtz instability. The size of the vortices is of the order of few Mm, which matches the transverse length scale in the Alfvén speed variation (Fig. \ref{fig:vortices_va}), as predicted by the straight flux-tube models \citep{antolin_2019}. The Kelvin Helmholtz instability in magnetised plasmas is inhibited by the magnetic tension, in this case however, the direction of the magnetic field is perpendicular to the flow velocity vector due to the oscillatory motion. The vortices can therefore develop without distorting the magnetic field, which would lead to stabilizing magnetic tension \citep{hillier_2019, barbulescu_2019}. We note that both presence of vortices and shear flows will contribute to the increased values of $\omega_x$ in the analysed region.

\section{Discussion}
\label{section:discussion}
\subsection{Oscillation parameters}
Displacement and velocity amplitudes in both instances of bundle oscillations are in agreement with the values typically observed in active region coronal loops \citep[e.g.][]{white_2012}, although this also depends on the type and magnitude of the perturbation responsible for the excitation of the oscillation. The detected oscillation periods are also within the range commonly seen in the observations \citep{nechaeva_2019}. We note that the oscillation periods decrease over the duration of the oscillation in both cases. This can be explained by changing physical properties in the magnetic loop bundle, such as changing loop length or distribution of the plasma along the loop. Such evolution is not surprising in a dynamic environment like this, as the length of the oscillating loops as well as the values of Alfvén speed can change on timescales comparable to the duration of the oscillation. The expected values of the oscillation periods estimated from $P \sim 2L/v_{\mathrm{A}}$ using the values for the average length and the Alfvén speed at the apex of the oscillating bundle loops at the beginning of each oscillation are 205 s and 175 s for oscillation 1 and 2 respectively. This roughly agrees with the detected initial periods, the change in the loop properties during the oscillation duration will however affect the oscillation periods. Changes in oscillation period over the duration of the oscillation due to changes in properties of the oscillating loop have both been seen in observations and been reproduced by numerical modelling \citep{kohutova_2017, verwichte_2017, su_2018}. Such shift in oscillation properties has a diagnostic potential using coronal seismology methods. A more detailed test of coronal seismology methods in convection-zone-to-corona simulations involving synthetic observables will be addressed in a follow-up study. Oscillation damping times are comparable to oscillation periods in both cases and in agreement with the observations of damped transverse oscillations, which are typically observed to decay within 3-4 oscillation periods \citep{goddard_2017}. We do not analyse the detailed profile of the oscillation damping, that is whether the damping profile is better represented by an exponential profile, Gaussian profile or a transition from Gaussian to an exponential profile as described in \citet{hood_2013, pascoe_2016}. Such a model is based on an assumption of a monolithic cylindrical coronal loop with a non-homogeneous boundary layer at the loop interface, which is not representative of coronal structures in the self-consistent convection-zone-to-corona simulations.

\subsection{Collective behaviour}
The bundle of magnetic loops in the simulation does not oscillate in isolation. It is in fact difficult to isolate the oscillating structures from a static, background plasma, as there is a large degree of collective behaviour among the structures in the simulated corona. The evolution of the individual magnetic loops in the bundle is not identical, but averaging the physical quantities at the bundle apex gives us the overall evolution of the bundle. This implies that care should be taken when modelling individual coronal loops as isolated structures. As the coronal loop evolution is coupled to the environment, the arcade model described by \citet{hindman_2021} is more representative of the evolution of the coronal structures in our simulation. The observations also suggest that the oscillations of individual coronal loops are often coupled to the oscillation of nearby magnetic structures and rarely occur in isolation \citep{verwichte_2004, verwichte_2009, jain_2015}. 

Regardless of the geometry, oscillating large scale structures create shear flows that lead to Kelvin-Helmholtz instability and the subsequent development of vortices if the instability is not inhibited by the magnetic tension. The shear created by large-scale translational motions also contributes to the development of Kelvin-Helmholtz instability, provided the direction of the motion is perpendicular to the coronal magnetic field \citep{hillier_2019, barbulescu_2019}. The development of vortices in our simulation may therefore be a result of a complex interplay of oscillatory and translational motions of coronal structures. The traditional models of Kelvin-Helmholtz unstable loops \citep[e.g.][]{terradas_2008, antolin_2014, karampelas_2017} assuming well-defined oscillating cylindrical loops embedded in a static plasma are however too idealised, as there are no loops with clearly defined cross-sections in our simulation. Because of a lack of clear loop boundaries the vortices are instead developed at the locations of maximum velocity shear.

\subsection{Oscillation damping and dissipation}
We stress the importance of the oscillation damping time $\tau$ as this corresponds to the rate at which the wave energy is either converted into another wave mode or dissipated. Empirical scaling of the damping time with the loop oscillation period is commonly cited as an indirect evidence for resonant absorption being the primary mechanism responsible for loop oscillation damping (see e.g. review by \citet{nakariakov_2020}). It has however been argued that the use of such scaling laws for discriminating between different damping mechanisms is questionable due to the inherent dependence of the damping time on the loop parameters \citep{arregui_2008}. We note that traditionally used scaling laws for resonant absorption assume large density contrast between the loop and the surrounding plasma \citep{ofman_2002}, which is not the case in our simulation, as shown in Fig. \ref{fig:context_full}. Oscillation damping times are further expected to depend on the magnetic Reynolds number given by $R_m = UL/\eta$, where U is a typical velocity scale, L is a typical length scale and $\eta$ is magnetic diffusivity. $R_m$ varies in the simulation as Bifrost uses $\eta(\vec{r},t)$ that is spatially and temporally dependent; $R_m$ in the simulation is however several orders of magnitude smaller than the estimated values in the solar corona.

We note that the traditional models for oscillation damping due to resonant absorption rely on a presence of an inhomogeneous layer at the boundary of a thin cylindrical loop with an Alfvén speed gradient \citep{ruderman_2002}. This is not a valid approximation for the loops in our simulation, event though Alfvén speed gradients are abundant in the simulated corona. We see no clear evidence of mode conversion in the oscillating bundle (that is, conversion to clearly identifiable spatially localised azimuthal oscillations); instead, the evolution of the physical quantities at the apex of the loop bundle suggests that the wave energy is dissipated into heat through the development of shear flows, leading to an increase in the viscous and resistive dissipation over the duration of the oscillation (in MHD models, the resistive and viscous dissipation terms are a parametrisation of processes operating on kinetic scales). Velocity shear leads to development of Kelvin-Helmholtz vortices which are detectable in the velocity field in the plane perpendicular to the bundle cross-section. The increase of the $\omega_x$ component in the vicinity of the bundle apex during the first oscillation decay suggests that the shear due to the oscillatory motion drives the development of vortices which then dissipate, leading to subsequent $\omega_x$ decrease effectively explaining oscillation damping and dissipation. However, the absence of such a clear vorticity peak at the loop apex during the second oscillation decay suggests the picture in this case is not as clear and multiple processes can contribute to the oscillation damping. The evolution of loop oscillations might be further affected by the motion of the loop footpoints, as these are not static. We note that the wave dissipation is only one of the several possible mechanisms contributing to the large temperature increase seen at the apex of the oscillating loops.

The onset of Kelvin-Helmholtz instability in models of transversely oscillating loops is linked to the development of turbulence leading to the formation of small scales which allows for fast dissipation of the wave energy \citep{hillier_2020}. For a loop to be considered truly 'turbulent' it is however necessary to demonstrate a non-linear cascade of energy to small scales. We cannot draw any conclusions about the presence of turbulent behaviour/lack thereof in our simulation due to limits posed by spatial resolution as well as by the magnetic diffusivity and viscosity. We also note that low Reynolds numbers in self-consistent MHD simulations artificially restrict the cascade to small scales \citep{howson_2017}. 

Kelvin-Helmholtz vortices developed during oscillation damping have also been proposed as being responsible for coronal oscillations appearing as decayless in certain emission lines \citep{antolin_2016}. In such model, however, the KHI vortices were formed at much smaller spatial scales due to the sharp density (and Alfvén speed) contrast at the loop boundary which is not the case for the loop bundle analysed in this work. 

\subsection{Implication for coronal loop models}
We find that oscillation parameters and evolution observed in coronal loops are reproduced by self-consistent simulations which include complex magnetic field geometry and density structuring and which do not contain well-defined coronal loops. This approach provides insight into how does the oscillating loop bundle actually evolve in three dimensions, including the detailed evolution of the magnetic field, into the degree of collective oscillation of the surrounding plasma and into the physical mechanisms associated with the oscillation damping and dissipation. This type of dynamics is impossible to capture by idealised straight flux tube models. This might have widespread implications for the accuracy of the coronal seismology methods, which are mostly based on cylinder approximations for coronal loops. Even recent numerical studies of the evolution of initially homogeneous coronal loops in response to transverse motions suggest that the highly idealised picture of coronal loops as monolithic plasma cylinders is unlikely to be realistic in the first place \citep{antolin_2019}. The question remains how realistic our current self-consistent simulations really are when it comes to reproducing detailed characteristics of the solar corona. However, despite the complex collective behaviour in such self-consistent simulations some results from simple coronal loop models are reproduced, namely the generation of the Kelvin-Helmholtz vortices by the transverse motions. The simple coronal loop models provide a lot of value for understanding fundamental physical processes at play. Caution should however be taken when drawing conclusions from observations whether the assumptions made in the models are still applicable in the analysed scenario. Combination of simple and self-consistent models is necessary for detailed understanding of the oscillatory behaviour in the corona.

Finally, we note that the length of loops we can simulate in this type of setup is limited by the size of the simulation domain, with the maximum length of magnetic loops in the simulation used in this work being 20 - 30 Mm. This affects the parameter space that is accessible by such models, as several oscillation parameters scale with loop length. Larger domains are therefore needed for more accurate one-to-one comparison of oscillation parameters with observations.

\section{Conclusions}
\label{section:conclusions}
For the first time, we analysed the damping of coronal oscillations using a self-consistent 3D radiation-MHD model of the solar atmosphere spanning from the convection zone into the corona, the associated oscillation dissipation and heating, and finally the physical processes responsible for the damping and dissipation. The simulated corona formed in such a model does not depend on any prior assumptions about the shape of the coronal loops. Using magnetic field tracing we analysed the evolution of a bundle of magnetic loops in the centre of the simulation domain. The magnetic bundle shows dynamic evolution and a large degree of collective behaviour of the individual loops in the bundle. We find that the bundle of magnetic loops shows damped transverse oscillations in response to perturbations in two separate instances with oscillation periods of 177 s and 191 s, velocity amplitudes of 10 km s$^{-1}$ and 16 km s$^{-1}$ and damping times of 176 s and 198 s, respectively. The oscillation parameters and evolution observed are in line with the values typically seen in observations of coronal loop oscillations. The oscillation periods decrease in both instances during the oscillations. We find that transverse coronal oscillations lead to the development of velocity shear in the simulated corona resulting in the formation of vortices with sizes around one Mm seen in the velocity field caused by the Kelvin-Helmholtz instability, contributing to the damping and dissipation of the transverse oscillations into heat. Assuming the structure of the corona in the self-consistent models is indeed realistic, our models of monolithic and static coronal loops with constant lengths might need to be reevaluated in favour of more realistic models accounting for loop properties changing with time and relaxing the assumption of highly idealised waveguides.

\begin{acknowledgements}
PK acknowledges funding from the Research Council of Norway, project no. 324523. This research was also supported by the Research Council of Norway through its Centres of Excellence scheme, project no. 262622 and through grants of computing time from the Programme for Supercomputing. PA acknowledges funding from his STFC Ernest Rutherford Fellowship (No. ST/R004285/2). 
\end{acknowledgements} 

\bibliography{waves}

\end{document}